\documentstyle[12pt]{article}
\def\({\left (}
\def\){\right )}
\def\[{\left [}
\def\]{\right ]}
\def\be{\begin{equation}}
\def\ee{\end{equation}}
\begin{document}
 \begin{titlepage}
 \vspace*{-4ex}
 \null \hfill MPI-Ph/92-41\\
 \null \hfill April, 1992 \\
\vskip 2.6 true cm
 \begin{center}
 {\bf \LARGE Finite Temperature Scalar Potential \\
       from a $1/N$ Expansion
    }\\[12ex]
{\large
  Vidyut Jain }
%$^1$
   \\ [6ex]
%

%$^1$
Max-Planck-Institut f\"{u}r Physik und Astrophysik \\
  - Werner-Heisenberg-Institut f\"{u}r Physik \\
 P.O.\,Box 40 12 12 , D - 8000 Munich 40, Germany \\ [2ex]
 \end{center}
\vskip 2.7cm
 \begin{abstract}
We compute the leading and next--to--leading corrections to
the finite temperature scalar potential for a
3+1 dimensional $\phi^4$ theory
using a systematic $1/N$ expansion. Our approach automatically
avoids problems associated with infrared divergences in
ordinary perturbation theory in $\hbar$. The leading
order result does not admit a first order phase transition.
The subleading result shows that the exact theory
can admit at best only a very weak first order
phase transition. For $N=4$ and weak scalar coupling we find
that $T_1$, the temperature at which tunneling from the origin
may begin in the case of a first order transition,
must be less than about 0.5 percent larger than $T_2$, the
temperature at which the origin changes from being a local minimum
to being a local maximum. We compare our results to the effective
potential found from a sum of daisy graphs.
 \end{abstract}
 \end{titlepage}

\noindent {\Large \bf 1. Introduction.}

There has been much recent interest in the nature of the electroweak
phase transition motivated by the possibility of baryogenesis within
the standard model itself. As noted by Kirzhnits amd Linde [1],
Weinberg [2] and Dolan and Jackiw [3] over 18 years ago, a spontaneously
broken field theory may have its symmetry restored at high enough
temperature. For example, in a spontaneously broken $\phi^4$ theory,
temperature corrections give a positive mass--squared contribution at the
origin ($\phi=0$), and at high enough temperature this correction
results in a global symmetry--unbroken minimum at the origin.

In such a model there are two important temperatures as we lower the
temperature from a very high value. The first, $T_1$, is the
temperature at which a second possible minimum appears degenerate in
energy with the minimum at the origin. The second, $T_2$, is the
temperature at which the effective mass at the origin vanishes, i.e.
when the origin changes from being a local minimum to a local
maximum.

The phase transition from the symmetric to nonsymmetric phase
as we cool a system described by such a model may therefore
proceed in two ways, by tunneling when the temperature is between
$T_1$ and $T_2$, or by a rollover when the
temperature drops below $T_2$. At $T=T_1$, isolated
bubbles of the symmetry broken phase will start to be created by
tunneling. If the phase transition completes by bubble nucleation
we will call the transition first order, otherwise if it completes mostly
by a rollover after $T<T_2$ then we will call the transition second
order. If the finite temperature
scalar potential possesses only the local minimum at $\phi^2=0$ until
$T=T_2$ then there is no $T_1$ and the phase transition is necessarily
second order. Even if there is a $T_1$ we will say only that the
system admits a first order transition since if $T_1$ is close
enough to $T_2$ we expect the phase transition to complete by a
rollover.

In the context of the standard model, it has been suggested that
sufficient baryogenesis may occur [4,5,6] if the phase transition
is first  order. Unfortunately, an accurate determination of
the the nature of the phase transition has proven difficult
in this case. One needs a reliable determination of the temperature
dependent effective potential of the scalars near the origin
for temperatures between $T_1$ and $T_2$. One--loop calculations
suggest that baryogenesis can only occur in the minimal standard
model for Higgs mass $M_H < 45\; GeV$ [7,8], a possibility that
appears ruled out by experiment [9]. However, as noted by both
Weinberg [2] and Dolan and Jackiw [3], naive finite temperature
perturbation theory for even the four dimensional
$\phi^4$ model suffers from infrared
divergences. These divergences invalidate the one--loop approximation.
In particular, at temperatures close to $T_2$ perturbation theory
in $\hbar$ begins to break down. The reason for this can be easily
understood:
even the one--loop correction drastically modifies the
tree level potential; it is clearly not a small perturbation.
Due to such problems, the one--loop result in pure $\phi^4$ theory
gives a complex potential for small values of $\phi^2$ and cannot
be used to study the nature of the phase transition.

In the standard model,
one may argue that if the phase transition completes ``well before''
$T_2$ then the one--loop calculation can be used. However, in the
case of the standard model the phase transition is expected to be
weakly first order, if it is first order at all. In addition, it is
not $a$ $priori$ clear that resumming some infinite class of diagrams
does not rule out a first order transition. The task then is to
extract the leading corrections
near $T=T_2$ to all orders in perturbation
theory. For example,
Dolan and Jackiw summed a class of diagrams, the ``super--daisies''
to circumvent the infrared divergence
problem in pure $\phi^4$ theory (with $N$ scalars)
to get a reliable estimate of $T_2$.
Such diagrams correspond to iterating the daisy graphs

\begin{picture}(400, 40)(0,20)
     \thicklines
\multiput(20,30)(80,0){5}{\circle{20}}
\put(120,30){\circle{20}}
\multiput(180,13)(0,34){2}{\circle{14}}
\put(260,47){\circle{14}}
\put(248,18){\circle{14}}\put(272,18){\circle{14}}
\put(328,18){\circle{14}}\put(352,18){\circle{14}}
\put(328,42){\circle{14}}\put(352,42){\circle{14}}
\put(375,27){$\cdots$}
\end{picture}
\be \ee
so that each petal has an arbitrary number of petals and so on.
Such a summation corresponds to the leading term in a
$1/N$ expansion and greatly modifies the 1--loop estimate.

In this article we will use the systematic $1/N$ expansion
method to calculate the leading and next--to--leading corrections
to the finite temperature effective scalar potential near the origin
in a weakly coupled $\phi^4$ theory in four dimensions.
In the case of the standard
model, one also needs to incorporate corrections to the scalar
potential from gauge loops. We do not address this problem here,
but will later comment on the applicability of our methodology
to a gauged $\phi^4$ theory.
In section two we review the $1/N$
expansion for a $\phi^4$ theory and present our computations.
In section three we discuss our results. Qualitatively, we
find that the leading order in $N$ correction does not admit
a first order transition. The next--to--leading may admit a first
order transition, but an  extremely weak one.
We stress that this approach automatically avoids the infrared
divergence problem of ordinary perturbation theory. In particular,
for the range of $\phi^2$ and $T$ we are interested in the potential
does not become complex.

In the context of the  standard model there have
been several recent papers on improving the 1--loop estimate [10,11,12].
These attempts mainly involve summing the daisy diagrams.
The simplest procedure [13] has been to use the 1--loop generated
temperature dependent mass to calculate 1--loop corrections. In pure
$\phi^4$ theory, if properly done, this can be interpreted as
summing the daisy diagrams.
This
procedure has its roots in the work of Weinberg [2]. It
is useful to briefly discuss this procedure, and at the
same time introduce some necessary formulas, before presenting our
results.

To understand the meaning of
such a procedure, it is necessary to recall an
important formula in the background field method. To be precise we
consider a simple $O(N)$ scalar field theory with action $(i=1\ldots N)$
\be
  S[\phi] = {1\over2}\int \delta^{ij}\partial_\mu \phi^i
     \partial^\mu \phi^j - {\lambda\over 4!}\int (\phi^2-v^2)^2. \ee
One calculates [14]
\def\hphi{\hat{\phi}}
\begin{eqnarray} \Gamma[\phi] &=& -i\hbar \ln \int [D\hphi]
   \exp \( {i\over\hbar}(S[\phi+\hphi]-{\delta S[\phi]\over
      \delta\phi^i}\hphi^i)\)\nonumber \\
  &=& S[\phi]-i\hbar\ln\int[D\hphi]\exp \({i\over\hbar}\int
    \(-{1\over2}\hphi^i\Delta^{-1}_{ij}\hphi^j-{\lambda\over 3!}
      \phi_k\hphi^k\hphi^2-{\lambda\over 4!}(\hphi^2)^2 \)\).
\end{eqnarray}
The first term on the RHS of the last line is the classical action.
The inverse scalar propogator is
\be \Delta^{-1}_{ij} = [\delta_{ij}\partial^2+M_{ij}^2(\phi)]
    . \ee
$M_{ij}^2$ is the second derivative of the classical potential
w.r.t. the fields $\phi^i$, $M_{ij}^2=\partial_i\partial_j V=
(\lambda/6)(\delta_{ij}(\phi^2-v^2)+2\phi_i\phi_j)$.

Equation (3)
is an effective action in that it incorporates the effects of all
one--particle--irreducible (1PI) diagrams and can be given to any
order in perturbation theory. It may be evaluated in perturbation
theory as
\def\Tr{{\rm Tr}}
\be
\Gamma[\phi] =
   S[\phi]+{1\over2}i\hbar\Tr\ln(\delta^{ik}\Delta^{-1}_{kj})
        + \bar{\Gamma}[\phi],
\ee
The
second term is the familiar one--loop contribution and the rest
contains all higher loop corrections from 1PI graphs.
The one-loop corrected potential is
\be \tilde{V}(\phi)
    = V(\phi)-{1\over2}i\int_p \Tr\ln \[p^2\delta_{ij}-M^2_{ij}\],
\ee
where the trace is over internal indices
and for zero temperature
the integration measure over momentum is $d^4p/(2\pi)^4$. For finite
$T$, in the imaginary time formalism,
the integral over four momentum goes over to an integral over
three momentum and a discrete sum in the manner given in [3]. We will
assume the reader is familiar with finite temperature field theory.

We now return to problem of understanding what it means to use the
one--loop corrected mass to calculate one--loop corrections.
If we iterate the one--loop result once we obtain
\be
 \tilde{\tilde{V}}(\phi) = V(\phi)-{1\over2}i\int_p
  \Tr\ln \[p^2\delta_{ij}-M^2_{ij}+{1\over2}i\int_k \partial_i\partial_j
         \Tr\ln\[k^2-M^2_{lk}\] \].
\ee
If we keep only the contributions from (7) when the $\partial_i
\partial_j$ both act on the same mass--squared matrix we obtain,
by carrying out the differentiation and expanding the log,
\be
  \tilde{\tilde{V}}=\tilde{V}+{i\over 2}\sum_{n=1}^\infty
     \Tr\int_p {1\over n} \[ i [p^2-M^2]_{mj}^{-1} \int_k
     {\lambda\over 6}(\delta_{ij}\delta_{kl}+\delta_{ik}\delta_{jl}
       +\delta_{il}\delta_{jk})[k^2-M^2]_{kl}^{-1} \]^n.
\ee
Here the term inside the square brackets is understood as a matrix
with two indices; the trace is over powers of this matrix.
It may be verified that the RHS is the tree--level plus
contributions from the daisy graphs of (1), with the correct
combinatorics $except$ for the two loop graph (n=1) which is
over counted. The correct factor for the two--loop graph is half
the one appearing from this procedure [14]. In the large $N$ limit
if we keep only the leading order in $N$ and account for the two
loop difference, this will reproduce the result of Dolan and
Jackiw [3] for the finite temperature daisy sum correction to the
effective mass at the origin:
\be
 {\partial \tilde{\tilde{V}}\over\partial\phi^2} =
 {\partial \tilde{V}\over\partial\phi^2} + {iN\over2}
   \sum_{n=1}^\infty \int_p {1\over n} \(\int_k [k^2-{\lambda\over 6}
     (\phi^2-v^2)]^{-1}\)^n {\partial \over\partial \phi^2}
     \( {iN\lambda\over6}[p^2-{\lambda\over 6}(\phi^2-v^2)]^{-1}\)^n.
\ee
Following [3], one keeps only the most infrared divergent contributions.
This amounts to letting the $\partial/\partial\phi^2 $ act only on
$[p^2-{\lambda\over6}(\phi^2-v^2)]^{-n}$ for $n\geq 2$. For the two
loop case it may act on either
$[p^2-{\lambda\over6}(\phi^2-v^2)]^{-1}$ or
$[k^2-{\lambda\over6}(\phi^2-v^2)]^{-1}$
which again explains why this case
is different. As shown in [3], however, the daisy graphs do not exhaust
all dominant $N$ contributions. For this one needs to sum the
superdaisy graphs. Therefore, even for large $N$, this approximation
scheme is not complete. Furthermore, in (7) there are additional
corrections when the $\partial_i$ and $\partial_j$ both act on different
mass--squared matrices. These corrections cannot be written as a
sum of 1PI diagrams. Therefore a sensible approximation scheme can
only be achieved by ignoring these corrections (which should in
any case be subleading in $N$). In the last
section we will compare our  $1/N$ expansion result with the
leading $O(N)$ correction from using an effective temperature
dependent mass.

\vskip 1.5cm

\noindent {\Large \bf 2. $\phi^4$ theory to subleading order.}

The systematic $1/N$ expansion allows us to calculate (3) as
a perturbation in $1/N$ near $\phi=0$ [15,16]. Root [16] has
evaluated the leading and next--to--leading corrections to the
zero temperature scalar potential in 4,3,2 and 1 dimensions. The
procedure at finite temperature is very similar.

At very high temperature, our results for four dimensions
should be essentially those of a three dimensional euclidean field theory
with a dimensionful $\phi^4$ coupling. In three dimensions and zero
temperature, the leading $O(N)$ potential has long been known [15].
It has exactly the same form as the sum of finite temperature superdaisy
graphs that were computed by Dolan and Jackiw [3] for a four
dimensional $\phi^4$ theory.
An important point in our approach is that
the effect of introducing an auxiliary field
$\sigma$ is to shift the $\phi$ mass term and  as a result there are
no infrared divergences in this formalism.

To proceed, we first set
$\lambda= f/N$. The $1/N$ expansion assumes $f$ $is$ $fixed$
as $N$ increases, not $\lambda$. Then by introducing a dimension
two auxiliary field $\sigma$ we rewrite the action (2):
\begin{eqnarray} S[\phi,\sigma] &=& S[\phi] + {3N\over 2f}
   \(\sigma-{f\over 6N}(\phi^2-v^2)\)^2
   \nonumber  \\ & = &
   {1\over2}\int (\partial\phi)^2+  {3N\over 2f}\int \sigma^2
      -{1\over2} \int (\phi^2-v^2)\sigma.
\end{eqnarray}
The auxiliary field has eliminated the $\phi^4$ term;
the original form of (3) is easily recovered by use of the
equation of motion for $\sigma$.

To calculate the effective potential $V(\phi)$ one proceeds as
follows. First, using the background field method one computes the
effective potential as a function of backgrounds of $\phi$ and
$\sigma$. Then, the background of $\sigma$ is
eliminated by its equation of motion.

The systematic $1/N$ expansion is performed by expanding the
action $S[\phi,\sigma]$ about backgrounds $\phi$ and $\chi$ thus:
\be \phi^i \rightarrow \sqrt{N}\phi^i+\hphi^i, \qquad
     \sigma \rightarrow \chi+ {\hat{\sigma}\over\sqrt{N}}. \ee
The factors of $\sqrt{N}$ have been inserted with hindsight. In
particular, the rescaling of the background $\phi$ is necessary
in order to ensure that the effective potential is renormalizable
to each order in the $1/N$ expansion. This will be seen later
in the explicit calculations.

One now writes a formula similar to (3), with an integral
over $\hphi$ and $\hat{\sigma}$:
\be \Gamma[\phi,\chi]=S[\sqrt{N}\phi,\chi]-i\hbar\ln\int
   [D\hphi][D\hat{\sigma}]\exp\({i\over\hbar}\int
   \(-{1\over2}\hphi^i\Delta^{-1}_{ij}\hphi^j-
   {\hphi^2\hat{\sigma}\over2\sqrt{N}} -\phi^i\hphi^i\hat{\sigma}
     +{3 \hat{\sigma}^2 \over2f} \)\).\label{seo}
\ee
The factor of $\hbar$ is useful in order to
compare results in the $1/N$ expansion to those from the usual
perturbation theory in $\hbar$; for practical purposes we will
take $\hbar=1$.
All terms linear in the quantum fields have been discarded, and
we have defined
\be \Delta^{-1}_{ij} = \delta_{ij}(\partial^2+\chi). \ee

The integral over $\hphi$ is gaussian and can  (formally) be
performed exactly to give the leading
$O(N)$ contribution to the effective potential. The integral over
over $\hphi$ also produces $\hat{\sigma}$ terms which may be
expanded in a power series. It is known that the $\hat{\sigma}^2$
terms give rise to the next--to--leading corrections in the $1/N$
expansion, which is all we are interested in. Furthermore,
to calculate the effective potential we can take $\phi$ and $\chi$
as space--time constants, but not $\hat{\sigma}$. The integral
over $\hphi$ produces a kinetic term for $\hat{\sigma}$ which must
be carefully determined in order to evaluate the next--to--leading
corrections.

The $\hphi\hat{\sigma}$ term in (\ref{seo}) can be rewritten as a
term purely  in $\hat{\sigma}$ by shifting the variable
of integration $\hphi$. This can easily be acheived to
all orders in the $1/N$ expansion, but since we are only
interested in the leading and next--to--leading terms it suffices
to write the argument of the exponential in (\ref{seo}) as
\be {i\over\hbar}\int \( -{1\over2}\hphi^i\Delta^{-1}_{ij}\hphi^j
   -{1\over 2\sqrt{N}}\hphi^2\hat{\sigma}+{1\over2}\hat{\sigma}^2
   \phi_i\Delta^{ij}\phi_j + {3\over 2f}\hat{\sigma}^2 \). \ee

\def\dJX{{\delta\over\delta J_i(x)}}
Using  $\hphi^i(x)=-i\hbar\dJX \exp({i\over\hbar}\int J_k \hphi^k)$
evaluated at $J_i=0$, and then rescaling the source and field
$\hphi$ by $\hbar^{1\over2}$ we obtain for the integral over
$\hphi$ in (\ref{seo}),
\begin{eqnarray}
  {\Gamma}[\phi,\chi,\hat{\sigma}] &=&
     -i\hbar\ln\int [D\hphi] \exp\( {i\over\hbar}\int \(
       -{1\over2}\hphi^i\Delta^{-1}_{ij}\hphi^j - {1\over2\sqrt{N}}
     \hphi^2\hat{\sigma}\)\) \nonumber \\
    &=& {1\over2}
  i\hbar\Tr\ln{\Delta}^{-1}_{ij} -i\hbar\ln\[ \exp\( {i\over\hbar}
  S_I[-i\hbar^{1\over2}{\delta\over\delta J}]\)
  \exp\(-{i\over2}\int J_i{\Delta}^{ij} J_j\)\],
    \label{pmr}
\end{eqnarray}
evaluated at $J=0$, where
\be S_I[\hphi] =
    -{1\over2\sqrt{N}}\int \hat{\sigma}\hphi^2. \ee

\def\ol{ \begin{picture}(22,12)(0,+8.5)
     \thicklines \put(11,11){\circle{20}} \end{picture} }
Of course, eq.  (\ref{pmr}) does nothing more than define a
perturbative expansion with a scalar propogator ${\Delta}^{ij}$.
The first term in (\ref{pmr}) is a one--loop result. It
has been evaluated many times before and is given by [3]
\begin{eqnarray} \ol & \ni & {i\over2}\hbar \int_{x,p} \Tr\ln\[
    p^2-\delta_{ij}\chi \] \nonumber \\
   & = & -\int_x V_0 + V_T .
\end{eqnarray}
$V_0$ is the zero temperature result which is divergent and must be
regulated. With a sharp momentum cutoff $\Lambda$,
\be V_0 = {N\over32\pi^2}\[{1\over2}\chi^2\ln[\chi/\Lambda^2]
      -{1\over4}\chi^2+\chi\Lambda^2 \].
\ee
$V_T$ is the finite, temperature dependent, result:
\be V_T = N\[ -{\pi^2\over90\beta^4}+{\chi\over 24\beta^2}
   -{\chi^{3\over2}\over12\pi\beta}-
     {1\over 64\pi^2}\chi^2 \ln[ \chi\beta^2]
   + {c\over 64\pi^2} \chi^2 \] .
\ee
We have dropped $O(\beta^2)$ terms and higher, and $c\approx 5.41$.
This result is valid at high enough temperatures. In the electroweak
model, it is known [8] that the high temperature expansion is
well justified at temperatures relevant for the  study of
the phase transition.

The $O(\hat{\sigma}^2)$ correction is a 1--loop contribution as well.
Extracting the 1PI part from (\ref{pmr}) we obtain to leading order
\be \ol \ni -{i\over4}\hbar\int_{x,p} {1\over p^2-\chi}
   \hat{\sigma} {1\over (p+i\partial)^2-\chi}\hat{\sigma}.
 \label{sp}
\ee
In deriving this result we took the functional derivatives in
(\ref{pmr}), fourier transformed the resulting $\delta$--functions
and dropped all total divergences. We further used properties of the
translation operators: $\exp(-ip\cdot x)H(-i\partial_p)\exp(ip\cdot x)
=H(x-i\partial_p)=\exp(-i\partial_x\cdot\partial_p) H(x)
  \exp(i\partial_x\cdot\partial_p)$. Dropping total divergences
assumes that the integral is well regulated (in three dimensions
it is finite). Eq. (\ref{sp}) may be evaluated using Feynman parameters;
the result is
\begin{eqnarray}
 \ol & \ni & -{i\over4}\hbar\int_x \hat{\sigma}\int_0^1 dz \int_p
    {1\over [ p^2-(\chi-z(z-1)\partial^2)]^2}\hat{\sigma}
    \nonumber \\
 & = & \hbar \int_0^1 dz \int_x {1\over 2^5\pi\beta} \hat{\sigma}
      {1\over\sqrt{\chi-z(z-1)\partial^2}}\hat{\sigma}
 \nonumber \\ & = &
 \hbar\int_x {1\over16\pi\beta} \hat{\sigma} {1\over\sqrt{+\partial^2}}
     \sin^{-1}\[ {1\over\sqrt{1+4\chi/\partial^2}} \] \hat{\sigma},
\end{eqnarray}
which is $1/\beta$ times the corresponding zero temperature result
in three dimensions [16]. The $p$ integral was evaluated using
$\int_p [p^2-\chi]^{-2}=-\partial^2/\partial\chi^2\int_p
  \ln[p^2-\chi]$  and only the leading temperature dependent
and field dependent correction was kept. The value of the arcsine
is uniquely determined since for $\partial^2=0$ the integral
over $z$ is unambiguous.

The final result for (\ref{seo}) to the required order is thus
\be \Gamma[\phi,\chi] = S[\sqrt{N}\phi,\chi] +
   {1\over 2}i\hbar\Tr\ln\Delta^{-1}_{ij} + \Gamma_1[\phi,\chi],
\ee
where the next--to--leading contribution from the gaussian
integral over $\hat{\sigma}$ is
\be \Gamma_1[\phi,\chi]= {1\over2}i\hbar\Tr\ln
   \[ 1+{\hbar f\over 24\pi\beta}{1\over\sqrt{+\partial^2}}\sin^{-1}
   \({1\over\sqrt{1+4\chi/ \partial^2}}\)
           + {f\over 3} \phi_i\Delta^{ij}\phi_j
    \]. \label{enl}
\ee
We now have all the formalism behind us. Before presenting
explicit calculations it is useful to rederive the leading
order ``mass--gap'' equation of [3] to see how the
infrared problem is automatically avoided.

To leading order in $N$ one drops the $\hat{\sigma}$ terms and eq.
(\ref{seo}) reduces to
\begin{eqnarray}
  \bar{\bar{\Gamma}}[\phi,\chi]&=& \int_x {N\over2}(\partial\phi)^2
     + {3N\over2f}\chi^2-{N\over2}\chi
    (\phi^2-v^2)+{1\over2}i\hbar\Tr\ln\Delta^{-1}_{ij}
     \nonumber \\ &=& N \int_x \[
   {3\over2f}\chi^2 -{1\over2}\chi(\phi^2-v^2)+{1\over2}i\hbar\int_p
   \Tr\ln[p^2-\chi] \].
\end{eqnarray}
For convenience, we have rescaled the $vev$, $v^2\rightarrow Nv^2$
with respect to the last section. The equation of motion for $\chi$,
\be {\partial \Gamma[\phi,\chi]\over\partial\chi}=0 \label{gez} \ee
gives the following equation
\be \chi={f\over6} (\phi^2-v^2)-i\hbar {f\over6}\int_p
   {1\over p^2-\chi} . \label{est} \ee
But $\chi$ is also just $-2\partial\bar{\bar{\Gamma}}/\partial\phi^2$,
which is the radiatively corrected ``effective mass'' to leading order
and is the reason why (\ref{est}) is called the ``mass--gap'' equation.
It is exactly the result derived in [3] by summing superdaisy graphs,
each of which is infrared divergent. The finite temperature, $T=
1/\beta$, result in four dimensions is [3,15,16]
\be \chi={f\over6} (\phi^2-v^2)+\hbar{f\over6}\({1\over 12\beta^2}
    - {\sqrt{\chi}\over4\pi\beta}\)+\ldots , \label{echi} \ee
where the ``$\ldots$'' refer to
\be
(c-{1\over2}-\ln[\Lambda^2 \beta^2])
   {\hbar f\over96\pi^2}\chi+ {\hbar f\Lambda^2\over
       96\pi^2}. \ee
This result is the sum of the $T$ dependent finite
part and the $T=0$ divergent part. The divergences may be
absorbed by renormalizing $v^2$ and $f$ and by introducing an
arbitrary mass scale $M$ [15]. Then for $f\sim O(1)$, and for all
reasonable values of $M\beta$ we can forget the extra terms at
high enough temperatures.

To compare, the zero temperature result for the action (2) in
three dimensions is [15]
\be \chi={f\over6} (\phi^2-v^2)-\hbar{f\over6}
     {\sqrt{\chi}\over4\pi}. \ee

Eq. (\ref{echi}) determines $T_2$, the temperature when the
effective mass vanishes at the origin, to leading order in $N$.
Putting $\chi=0$ we get ($\hbar=1$)
\be {1\over \beta_2^2} = 12v^2 .\ee
Furthermore, at the origin, $\sqrt{\chi}$ has a remarkably
simple solution for $T$ just above $T_2$ [3]
\be \sqrt{\chi} = {2\pi\over 3}\( {1\over\beta}-{1\over\beta_2}\)
   .  \label{cjct} \ee
In general, we find for $\sqrt{\chi}$ the leading result
\be \sqrt{\chi} = {f\over 48\pi\beta} \[
    \sqrt{1+{32\pi^2\over f}(12\phi^2\beta^2-12v^2\beta^2+1)}-1\].
    \label{schi}
\ee
We have chosen the sign in the solution (\ref{schi}) of (\ref{echi})
so that $\sqrt{\chi}$ is positive for $T$ bigger than $T_2$. Eq.
(\ref{schi}) simplifies in various limits. At $\phi^2=0$, for $T$
just above $T_2$ we obtain (\ref{cjct}).
At $384\pi^2\phi^2 \beta_2^2 \gg f$ and $T$ just above $T_2$,
\be \sqrt{\chi} \approx \sqrt{{f\phi^2\over 6}}. \label{scap} \ee
Finally, when $T/T_2-1\gg f/(64\pi^2)$ (i.e. $32\pi^2(1-12v^2\beta^2)/f
\gg 1$) then for all $\phi$ we have the simplification
\be \sqrt{\phi}\approx \sqrt{{f\over6}} \sqrt{{1\over12}\(
    (T-T_2)^2+2(T-T_2) T_2\) +\phi^2}. \ee

We now return to the computation of $\Gamma_1$ in the high
temperature limit. Some care must be taken to ensure that the
final answer contains no temperature dependent divergences [17].
Also, to the order we are working in, it is sufficient to use the
solution of (\ref{echi}) for $\chi$ [16]. We will drop all
field independent divergences.

We make the replacement
$\partial^2\rightarrow -p^2$ and the trace becomes an integral
over space--time and momentum. At finite $T$,
$i\int_p\rightarrow -\beta^{-1}\int
 d^3\vec{p}^2/(2\pi)^3 \sum_n = -[4\pi^2\beta]^{-1}\int_0^\infty
 (d\vec{p}^2)\sqrt{\vec{p}^2} \sum_n $
with $-p^2\rightarrow \vec{p}^2 + 4\pi^2 n^2 T^2$ in the
integrand. The discrete sum over $n$ runs over all integers.
One then obtains,
\begin{eqnarray}
 \Gamma_1 &=& -{\hbar\over2\beta}\int {d^3\vec{p}^2\over (2\pi)^3}
  \sum_{n=-\infty}^{+\infty}
\nonumber \\ & & \!\!\!\!\!\!\!\!\!\!\!\!\!\!\!\!\!\!\!
 \ln\[ 1 + {\hbar f/
    (24\pi\beta)\over \sqrt{\vec{p}^2+4\pi^2 n^2 T^2} } \sin^{-1}\(
    {1\over \sqrt{1+4\chi/(\vec{p}^2+4\pi^2 n^2 T^2)} } \)
  + {f\phi^2/3\over \vec{p}^2+4\pi^2 n^2 T^2 +\chi} \] \!\! .
    \label{enlb}  \end{eqnarray}

The integral is not infrared divergent, but is ultraviolet divergent.
The ultraviolet divergences should be the same as those appearing
if we did the momentum integral over four momenta. [This not quite
the same as taking $\beta\rightarrow \infty$ here, because we use the
high temperature result for the $\hat{\sigma}$ propogator.] We must
isolate these divergences before calculating the temperature
dependent part in the high temperature limit. Furthermore, even for
say the $n=0$ contribution
 we were not able to find an analytic expression
for the integral. We can however evaluate it in some limits.
[For the rest of this section, we delete overall integrals over
space--time.]

Consider the $n=0$ contribution.
We then change variables: $t=\sqrt{4\chi/\vec{p}^2}$.
Equation (\ref{enlb}) becomes
\be \Gamma_1(n=0) = -{(4\chi)^{3\over2}\over (2\pi)^2\beta}\hbar
   \int_0^\infty {dt\over t^4}\ln\[1+\alpha t\tan^{-1}\({1\over t}\)
  + {f\phi^2\over3\chi} {t^2\over t^2+4}\],
   \label{tgr}
\ee
where we have defined $\alpha=\hbar f/(24\pi\beta\sqrt{4\chi})$ and
used the identity $\sin^{-1} (1+t^2)^{-1/2}=\tan^{-1} t^{-1}$.
At $\phi^2$=0,
$\alpha$ is small to temperatures just above $T_2$ because
from (\ref{cjct}) we have $\alpha<1$ as long as $1-\beta/\beta_2
  > f/(32\pi^2)$ at the origin. In what follows, we will assume
small $\alpha$ and small enough
$\phi^2/\chi$.
In (\ref{tgr}),
as $t$ ranges from 0 to $\infty$, $t\tan^{-1}t^{-1}$ ranges from
0 to 1. At $t=1$, $t\tan^{-1}t^{-1}=\pi/4$. Therefore for small $\alpha$,
$\alpha t\tan^{-1}t^{-1}$ remains small for all values of $t$ in
the integral. Hence,
 for small $\alpha$ and small $\phi^2$ we can use the
simplification $\ln[1+r]\approx r$ to simplify (\ref{tgr}) or
(\ref{enlb}).

\def\pp{\vec{p}^2+4\pi^2 n^2 T^2 }
For the moment let us ignore the explicit $\phi^2$ term in $\Gamma_1$.
To calculate (\ref{enlb}) for small $\alpha$
we note the following observation for the $n\neq 0$ contributions.
If $\sqrt{\chi} < \pi T$ then the arcsine possesses an
expansion in inverse powers of the momentum. We have,
\be  {\sin^{-1}\(
    {1\over \sqrt{1+4\chi/(\vec{p}^2+4\pi^2 n^2 T^2)} } \)\over
 \sqrt{\vec{p}^2+4\pi^2 n^2 T^2} }
    = {\pi/2\over \sqrt{\pp}}-\sum_{l=0}^{\infty}
       {(-1)^l (4\chi)^{l+{1\over2}}\over (2l+1)(\pp)^{l+1} } .
    \label{cex} \ee
The $n=0$ contribution unfortunately does not possess such a
straightforward expansion for all $\vec{p}^2$. If $\vec{p}^2>4\chi$
then the expansion is as above, otherwise it is
\be  {\sin^{-1}\(
    {1\over \sqrt{1+4\chi/\vec{p}^2} } \)\over
 \sqrt{\vec{p}^2} }
    = {1\over\sqrt{\chi}}\sum_{l=0}^{\infty}
       {(-1)^l (\vec{p}^2)^l\over (2l+1)\chi^l} .\label{sex}\ee
We then split up $\Gamma_1$ into four parts: $A$, the contribution
from $n=0$ for $\vec{p}^2>4\chi$; $B$, the contribution from
$n=0$ for $\vec{p}^2<4\chi$; $C$, the contribution from $n\neq 0$
for $\vec{p}^2> 4\chi$; $D$, the contribution from $n\neq 0$ for
$\vec{p}^2< 4\chi$. $B$ is found using the expansion (\ref{sex}), whereas
all the rest involve the expansion (\ref{cex}).
We denote the contributions
to $A$, $C$ and $D$ from the first term
in (\ref{cex}) by $a$, $c$ and $d$,
respectively. We further define $b$ to be a contribution similar to $a$
but with the momentum integral up to $\vec{p}^2< 4\chi$. Then it can be
seen $a+b+c+d$ is field independent and can be dropped, i.e. we evaluate
$\Gamma_1=(A-a)+B+(C-c)+(D-d)-b$.
For $B$ we get
\be B = -{(4\chi)^{3\over2}\over (2\pi)^2\beta}\hbar
   \int_1^\infty {dt\over t^3}\alpha \tan^{-1}\({1\over t}\)
     \approx -{f\hbar^2\chi\over 84\pi^3\beta^2}.
\ee
For $b$ we get
\be b={-\hbar^2 f\over 48\pi \beta^2}\int_0^{\vec{p}^2=4\chi}
   {d^3\vec{p}\over (2\pi)^3} {\pi\over 2\sqrt{\vec{p}^2}} =
      -{\hbar^2 f\chi \over 96\pi^2\beta^2}. \ee
We also get
\begin{eqnarray}
 (A-a)+(C-c) &=& {\hbar^2 f\over 48\pi\beta} \sum_{l=0}^{\infty}
   I_{l+1} {(-1)^l (4\chi)^{l+{1\over2}}\over (2l+1) }\nonumber \\
&\approx&
{\hbar^2 f\sqrt{\chi}\over 24\pi\beta}\( {\Lambda^2\over 16\pi^2}
     + {1\over 12\beta^2}\) - {3\hbar^2 f\chi\over 56\pi^3\beta^2}
       , \end{eqnarray}
where $I_l$ is given in the appendix (for $\epsilon=\sqrt{4\chi}$,
$\chi=0$). Furthermore it is straightforward to show that $D-d$ is
subleading in temperature and may be dropped.

The $\phi^2$ term from the expansion of the log is more easily
evaluated and contributes
\be \Gamma_1 \ni -{\hbar f\over 6} \phi^2 I_1 =
     -{\hbar f\over 6} \phi^2 \(
{\Lambda^2\over 16\pi^2}+{1\over 12\beta^2}-{\sqrt{\chi}\over 4\pi\beta}
     \), \ee
where $I_1$ is taken from the appendix ($\epsilon=0$, $\chi\neq 0$).
If we keep all powers of $\phi^2$, but ignore the arcsine term, we
must also include
\be \Gamma_1 \ni {\hbar \chi^{3\over2}\over 4\pi\beta}
     \sum_{l=2}^{\infty} {(-f\phi^2/6\chi)^l(2l-5)!!\over
        l!} . \ee
This series converges for all values of $\phi^2/\chi$ we will
be interested in.

Each of the $O(\Lambda^2)$ divergent terms in the above
equations is not separately renormalizable, however their sum
is. This is because for the next--to--leading terms we can use
eq. (\ref{echi}) to express  $\phi^2-\hbar\sqrt{\chi}/4\pi\beta
= 6\chi/f+v^2-\hbar(12\beta^2)^{-1}+O(1/N)$.
Since the tree-level potential
contains a term proportional to $v^2\chi$, we see that the
above divergences in $\Gamma_1$ can be absorbed by renormalizing
$v^2$ (throughout we have neglected field independent constants).
The dominant contributions to $\Gamma_1$ are found by summing
the above results; we obtain
\begin{eqnarray}
 \Gamma_1 &\approx& -\hbar\chi\({\Lambda^2\over 16\pi^2}
  +{1\over12\beta^2}\)+{\hbar f\phi^2\sqrt{\chi}\over 24\pi\beta}
   +{\hbar\chi^{3\over2}\over 4\pi\beta} \sum_{l=2}^{\infty}
   (-f\phi^2/6\chi)^l(2l-5)!!/l! , \nonumber \\
  &\approx& -\hbar\chi\({\Lambda^2\over 16\pi^2}
  +{1\over12\beta^2}\)+{\hbar \chi^{3\over2}\over 12\pi\beta}
     \[(1+f\phi^2/3\chi)^{3\over2}-1\]. \end{eqnarray}

Altogether, summing the various partial results we find for the
dominant high temperature effective scalar potential from (\ref{seo}),
$V(\phi)=-\Gamma[\phi,\chi(\phi)]$,
the following renormalized result up to a constant ($\hbar=1$):
\be
  V(\phi) = N\[ {1\over2}\chi(\phi^2-v^2)-{3\over2f}\chi^2\]
   + (N+2) {\chi\over 24\beta^2} - {\chi^{3\over2}\over 12\pi\beta}
    \[(1+f\phi^2/3\chi)^{3\over2}+N-1\].\label{mrr}
\ee
where $\chi$ is a solution of $\partial V/\partial\chi=0$.
We do not give details of the renormalization which for
zero temperature and in 3 and 4 dimensions may be found in [15,16].

We will use these results in the last section of this paper, where
we also discuss the range of $\phi^2$ and $T$ for which these
results are the most dominant.

\vskip 1.5 cm
\noindent{\Large\bf 3. Discussion.}

Our main result is the leading order in $N$ and next--to--leading
order expression for the high temperature scalar potential,
eq. (\ref{mrr}). The tree level potential, with the normalizations
of the last section is
\be V(\phi)= {Nf\over 4!} (\phi^2-v^2)^2. \label{vtree} \ee
Eq. (45) includes loop corrections to this.
We assumed $\beta\sqrt{\chi}\ll 1$ and
$\beta\sqrt{\chi}\gg f/48\pi \approx f/150$ in obtaining the
next--to--leading result. At $T=T_2$ for the leading order result,
this gives approximately
$100 v^2/f \gg \phi^2 \gg fv^2/100$. At $\phi^2=0$ we require
$T-T_2 \gg fT_2/300$.

To study the nature of the phase transition we may look for
zeros of $dV/d\phi^i$. There is always one zero at the origin,
$\phi^i=0$ for all $i$. Away from the origin we can look for
zeros of $dV/d\phi^2=(\partial V/\partial\chi)(\partial\chi/
\partial\phi^2)+\partial V/\partial\phi^2=\partial V/\partial\phi^2$.
Since to $O(N)$, $dV/d\phi^2\propto \chi$,
the leading order result has one such zero away from the
origin when $T<T_2$ (which is the local minimum at temperatures
below $T_2$). Thus, the leading order result does not admit
a first order phase transition, for which we need  zeros
away from the origin for temperatures $T \geq T_2$.

At next--to--leading order, the critical temperature $T_2$ is modified
from the leading order result. $\partial V/\partial\phi^2 =0$
for $V$ given by (\ref{mrr}) still has a solution when $\chi=0$.
Writing out $\partial V/\partial\chi$ at $\phi^2=0$ and setting
$\chi=0$ immediately gives for the temperature $T_2$ ,
\be
   \(1+{2\over N}\) T_2^2 = 12v^2. \label{ctr}
\ee
$T_2$ has been reduced from its leading order value.
$\partial V/\partial \phi^2=0$ has one further solution at the
origin. $\partial V/\partial\phi^2=0$ gives
\be {N\over 2}\sqrt{\chi}
   = {f\over 24\pi\beta} (1+f\phi^2/3\chi)^{3\over2}
  , \label{evpz} \ee
when $\sqrt{\chi}\neq0$. At $\phi^2=0$ this has the solution
$\sqrt{\chi}=f/(12N\pi\beta)$. Writing out $\partial V/\partial\chi=0$
and inserting the above expression for $\sqrt{\chi}$ gives
a second (slightly higher) temperature when the effective mass at
the origin vanishes. To $O(1/N)$ it is given by
\be \(1+{2-f/4\pi^2\over N}\) T^2= 12v^2 .\ee
For $N=4,f=1$ this is within 0.5 percent of $T_2$. Between this
temperature and $T_2$ the effective mass at the origin is negative!
However, since our approximations break down at $\beta\sqrt{\chi}$
less than about $f/48\pi$ such a phenomenon is not necessarily
physical; here it is an artifact of our approximations.

To ascertain the nature of the phase transition we have to study
the shape of the potential near $T=T_2$. If a second minimum
degenerate in energy with the one at the origin does not appear
by temperature $T_1>(1+5f\times 10^{-3})T_2$ then we cannot reliably
determine if the theory admits a first order phase transition.
However, in this case, if the exact theory does admit a first
order phase transition then it will be only a very weak one.

The general solution of $\chi(\phi)$ is complicated. As noted by
Root [16], one can use the leading order result for $\chi(\phi)$ in
the next--to--leading potential (\ref{mrr}) to find $V(\phi)$ to $O(1)$
in the $1/N$ expansion. However, (\ref{schi}) can develop an imaginary
part for temperatures below the leading order value for $T_2$. Therefore,
it is preferable to use the
next--to--leading order solution for $\chi(\phi)$.
Our results in what follows are correct to $O(1/N)$, barring the other
approximations we made.
At $T$ near $T_2$ given by
(\ref{ctr}) we once again have (\ref{scap}) when
$\phi^2\gg v^2/300$. To see this,
we write out the equation for $\chi(\phi)$,
\begin{eqnarray}
{6\chi\over f}& =& \phi^2-v^2+{1+2/N \over 12\beta^2}
-{\sqrt{\chi}\over4\pi\beta}
 \nonumber \\ & & -
{\sqrt{\chi}\over 4\pi\beta N}\[(1+f\phi^2/3\chi)^{3\over2}
 -1\] +{f\phi^2\over 12\pi\beta\sqrt{\chi} N}(1+f\phi^2/3\chi)^{1\over2}
   . \label{nec} \end{eqnarray}
In this put $\beta=\beta_2$ and
$\sqrt{\chi}=\sqrt{f\phi^2/6}+\delta$. Assuming
that $\delta$ is small one obtains
$\delta\approx -f/48\pi\beta [1+(\sqrt{3}
-1)/N]$ which can be ignored when
$\delta\ll \sqrt{f\phi^2/6}$. We can improve
on this estimate for $\sqrt{\chi}$ by noting that when $\sqrt{\chi}
\approx \sqrt{f\phi^2/6}$ the last
two terms in (\ref{nec}) add to an amount
which for large enough $N$ (say $N>3$)
are much smaller than the third last
term. Hence we have the approximate solution
\be \sqrt{\chi} = {f\over 48\pi\beta}\[\sqrt{1+{32\pi^2\over f}
      (12\phi^2\beta^2-12v^2\beta^2+1+2/N)}-1\]. \label{ass} \ee
This should be valid at temperatures near $T_2$ and $\phi^2$ sufficiently
far from the origin.

To see if a first order phase transition is possible we must find the
zeros of $dV/d\phi^2=\partial V/\partial\phi^2$.
There is one solution at $\sqrt{\chi}=0$ and another given by solving
(\ref{evpz}), which
 at $T=T_2$ and $\chi\approx f\phi^2/6$  has a solution at
$\sqrt{\phi^2} \approx v/\sqrt{{3f/[2N(N+2)]}}. $
This result holds if $N$ is not too large ($\sqrt{\phi^2}$ is not
too small). At $N=4$ we have $\sqrt{\phi^2}\approx \sqrt{f}v/4$.
We can do better numerically.
We studied the case $N=4$, $f=1$. The approximate solution
(\ref{ass}) suggests
$   T_1 = T_2 (1+3.5\times 10^{-3}), $
with $\sqrt{\phi^2}\approx 0.08v$ at the second minimum at $T=T_1$.
The height of the barrier separating the  two minima is at most
of $O(2\times 10^{-7} v^4)$. At $T=T_2$ the minimum is at
$\sqrt{\phi^2}\approx 0.15v$, which is slightly smaller than our cruder
estimate above.

All the numbers above occur as our approximations break down so
we cannot determine the exact nature of the phase transition. However,
as argued above we may deduce from them that the exact model can
admit only a very
weak first order phase transition. To get a more accurate picture
one should evaluate (35) in the limit $\beta\sqrt{\chi}\ll f/48\pi$.

To compare these results with the effective mass insertion
procedure discussed in the introduction, we note that the
leading $O(N)$ one--loop corrections to (\ref{vtree}) give
the (renormalized) potential [3]
\be V= N\[ {f\over4!} (\phi^2-v^2)^2 + {f(\phi^2-v^2)/6
  \over 24\beta^2}-{\(f(\phi^2-v^2)/6\)^{3\over2}\over 12\pi\beta}
    \] .\ee
As indicated in the introduction, this is complex for $\phi^2<v^2$.
The leading $O(N)$ one--loop corrected mass is
\be  {Nf\over 6} \(\phi^2-v^2+{1\over 12\beta^2}\). \ee
Using this to perform the one--loop corrections, one gets
\be V= N\[ {f\over4!} (\phi^2-v^2)^2 + {f(\phi^2-v^2+1/12\beta^2)/6
  \over 24\beta^2}
-{\(f(\phi^2-v^2+1/12\beta^2)/6\)^{3\over2}\over 12\pi\beta}
    \] .\ee
The effective mass at the origin vanishes at
$T_2^2=12v^2$, as for the leading
$O(N)$ result from the $1/N$ expansion. However, there is a
higher temperature at which this also occurs. Namely, if we put
$T=T_2+t$ then for small $t$ the appropriate equation is easy to solve.
The result is that the effective mass at the origin
also vanishes at $T\approx(1+f/144)T_2$. Between this temperature
and $T_2$ the effective mass at the origin is negative.
Thus compared to our results from the $1/N$ expansion, we see that
the effective mass insertion technique is not reliable
at temperatures less than  about 1 percent above  $T_2$ even in
the large $N$ limit.

In conclusion, for large enough $N$ the model can admit only a
very weak first order phase transition. For $N=4$ it would be interesting
to see how $O(1/N)$ contributions modify the results above. At $N=4$
the next--to--leading corrections are certainly smaller than the
$O(N)$ corrections and we do not expect $O(1/N)$ corrections to
compete with the $O(N)$ or $O(1)$ contributions. However, it would
be necessary to perform an explicit calculation to check this
belief. Finally, in the context of the standard model, one needs to
add gauge fields $A$ to this model. In fact, the contributions from
gauge loops are expected to dominate over those from scalar loops
in the weak scalar self--coupling limit. The main problem in
the case of a gauged model is the gauge fixing dependence of the
computations. Otherwise, in principle, is should be possible to
determine the dominant contributions to the finite temperature
scalar potential. For example, for $SU(2)$, the gauge kinetic
term has two types of $A^4$ interaction ($[A_\mu^i A^\mu_i]^2$
and $[A_\nu ^i A^\mu_i]^2$, where $i=1,2,3$) as well as
an $A^3$ interaction. Loops from the second $A^4$ interaction and
loops from the $A^3$ interaction are subleading to those from the
first $A^4$ interaction. If we drop these last two interactions
then what remains can be treated very similarly to the pure
$\phi^4$ theory. One  must then determine  what
gauge fixing invariant quantities can be extracted from such
a computation. It is expected that, if carefully done,
physically measurable quantities should be gauge fixing independent.
Work along these lines is in progress.

\vskip 1.0 cm
\noindent {\bf Acknowledgements.} \\
I am greatly indebted to G. Anderson, M.K. Gaillard, S. Naik and
especially P. Weisz, who introduced me to the $1/N$ expansion. I also
want to thank him, M. Carena and C. Wagner for useful comments
on the manuscript, and  H. Ewen, G. Lavrelashvili, W. M\"osle and
O. Ogievetsky for other useful discussions.

\newpage
\noindent{\Large\bf Appendix.}

The finite temperature quantities we need to evaluate in this
paper are all of the form
\be I_l =  {1\over \beta}\sum_{n=-\infty}^{+\infty} \int_{\vec{p}^2
   =\epsilon^2}^{\infty} {d^3\vec{p}\over (2\pi)^3}
   [\vec{p}^2+4\pi^2 n^2 T^2 +\chi ]^{-l} \ee
for $l$ a positive integer. The discrete sum runs
over all integer $n$. We need only the cases $\epsilon=0$ and $\chi
\neq 0$ or $\epsilon\neq 0$ and $\chi=0$.
We assume $\beta\sqrt{\chi}
\ll 1$, but $\beta\sqrt{\chi}\gg f/(48\pi)\approx f/150$.

For $l>1$ we can find the answer by
differentiating the $l=1$ result $l-1$ times w.r.t. $\chi$.
To evaluate the expressions we follow [3]. The sum over $n$ is
explicitly performed to give
\def\vp{\vec{p}}
\be
I_1 = \int_{\vec{p}^2=\epsilon^2}^{\infty} {d^3\vec{p}\over
    (2\pi)^3} \[{1\over 2\sqrt{\vp^2+\chi}}+{1\over
       \sqrt{\vp^2+\chi}\(\exp(\beta\sqrt{\vp^2+\chi})-1\)} \].
\ee
The first term in this expression
is the $T=0$ piece and contains all the ultraviolet
divergences; the second term is the $T$ dependent piece and is
finite. Changing variables, the $T$ dependent piece is
\be
 I_1(T\neq 0) = {1\over 2\pi^2\beta^2} \int^{\infty}_{\beta\epsilon}
    {z^2 dz\over \sqrt{z^2+\beta^2\chi}\(\exp(\sqrt{z^2+\beta^2\chi})
     -1\) }. \ee

In the high temperature limit, and for $\epsilon=0$, we
have the leading result of [3],
\be I_1(\epsilon=0) = {\Lambda^2\over 16\pi^2}+{\chi\over 8\pi^2}
    \ln\[{\sqrt{\chi}\over\Lambda}\] +
{1\over 12\beta^2}-{\sqrt{\chi}\over 4\pi\beta}
  + {\chi\over 16\pi^2}\[ c-{1\over2}-2 \ln[\beta\sqrt{\chi}]\]
  \label{djr} . \ee
The log pieces combine to give a single term proportional to
$\chi\ln[\beta\Lambda]$. Such log--divergent terms in the final
answer must be renormalizable. The coefficient of this term is
of $O(\beta\sqrt{\chi}/(2\pi))$
down from the preceeding $O(T)$ term,
and in our approximation all $O(\chi)$ terms can be neglected
in comparison to $O(\sqrt{\chi}/\beta)$ terms. Furthermore, to
properly renormalize such terms in the effective potential we
should keep subleading corrections to the $\hat{\sigma}$ propogator,
eq. (21), which we discarded. Therefore it is consistent to drop
all the terms proportional to $\chi$. This is what we do in the
following.

For $\epsilon\neq 0$ we subtract from the above the temperature
dependent piece
\begin{eqnarray}& &
 {1\over 2\pi^2\beta^2}\int_0^{\beta\epsilon} {z^2 dz\over
   z^2+\beta^2\chi}\[{\sqrt{z^2+\beta^2\chi}\over \exp(
     \sqrt{z^2+\beta^2\chi}) -1} \] \approx
{1\over 2\pi^2\beta^2}\int_0^{\beta\epsilon}
     {z^2 dz\over z^2+\beta^2\chi}\(1-{1\over2}\sqrt{z^2+\beta^2\chi}\)
\nonumber \\& & \qquad\qquad
    \approx  {\epsilon\over 2\pi^2\beta}-{\sqrt{\chi}\over 2\pi^2
       \beta}\tan^{-1}\[{\epsilon\over \sqrt{\chi} }\]
       - {\epsilon\sqrt{\epsilon^2+\chi}\over 8\pi^2}+
      {\chi\over 8\pi^2}\ln\[ {\sqrt{\epsilon^2+\chi}+\epsilon \over
          \sqrt{\chi} } \],
\end{eqnarray}
and the zero temperature piece,
\be {1\over 4\pi^2}\int _0^{\epsilon} {z^2 dz\over \sqrt{z^2+\chi}}
     = {\epsilon\sqrt{\epsilon^2+\chi}\over 8\pi^2}
           -{\chi\over 8\pi^2}\ln\[ {\sqrt{\epsilon^2+\chi}+\epsilon
        \over \sqrt{\chi}  } \], \ee
to arrive at
\be
I_1(\epsilon\neq 0) = {\Lambda^2\over 16\pi^2}+
{1\over 12\beta^2}-{\epsilon\over2\pi^2\beta}
+{\sqrt{\chi}\over2\pi^2\beta}
\(\tan^{-1}\[{\epsilon\over \sqrt{\chi} }\]
-{\pi\over2}\).\label{www}
 \ee
In the limit $\chi\rightarrow 0$ this expression has an expansion
in positive integer powers of $\chi$. Therefore, its $l$th derivative
w.r.t. $\chi$ evaluated at $\chi=0$ is well defined.
By expanding the arctan in this limit we obtain for $l$ a positive
integer and $\chi=0$,
\be
I_l(\epsilon\neq 0) =
    \[{\Lambda^2\over16\pi^2}+{1\over 12\beta^2}\]\delta_{l-1}
+{\epsilon^{3-2l}\over
      2\pi^2\beta (2l-3) }
         . \ee
$I_l$ for $\epsilon=0$ and $\chi\neq 0$ is simply found by
differentiating (\ref{www}) w.r.t. $\chi$ the appropriate number
of times. The answer is, for $l>1$,
\be I_l(\chi\neq 0) =
{\chi^{{3\over2}-l}\over 4\pi\beta }{(2l-5)!!\over 2^{l-1}(l-1)!}
    . \ee

\end{document}